\documentclass{article}
\usepackage[dvips]{graphicx}  %  XXX archive  version
\usepackage{color}

\usepackage[english]{babel} % Babel

\usepackage[varg]{txfonts}
\usepackage{concmath}
\begin{document}

\newcommand{\rum}{\rule{0.5pt}{0pt}}
\newcommand{\rub}{\rule{1pt}{0pt}}
\newcommand{\rim}{\rule{0.3pt}{0pt}}
\newcommand{\numtimes}{\mbox{\raisebox{1.5pt}{${\scriptscriptstyle \rum\times}$}}}
\newcommand{\numtimess}{\mbox{\raisebox{1.0pt}{${\scriptscriptstyle \rum\times}$}}}
\newcommand{\Boldsq}{\vbox{\hrule height 0.7pt
\hbox{\vrule width 0.7pt \phantom{\footnotesize T}%
\vrule width 0.7pt}\hrule height 0.7pt}}
\newcommand{\two}{$\raise.5ex\hbox{$\scriptstyle 1$}\kern-.1em/
\kern-.15em\lower.25ex\hbox{$\scriptstyle 2$}$}

\renewcommand{\refname}{References}
\renewcommand{\tablename}{\small Table}
\renewcommand{\figurename}{\small Fig.}
\renewcommand{\contentsname}{Contents}

\twocolumn[%
\begin{center}
\renewcommand{\baselinestretch}{0.93}
{\Large\bfseries Dynamical 3-Space: Black Holes in an Expanding Universe

}\par
\renewcommand{\baselinestretch}{1.0}
\bigskip
David P. Rothall$^*\!$ and  Reginald T. Cahill$^\dagger$\\ 
{\footnotesize  School of Chemical and Physical Sciences (CaPS), Flinders University, SA 5042, Australia\rule{0pt}{12pt}\\
$^*$E-mail: david.rothall@flinders.edu.au\\
$^\dagger $ Email: reg.cahill@flinders.edu.au \\

Progress in Physics  Vol 4, 25-31, 2013

}\par
\medskip
{\small\parbox{11cm}{%
Black holes are usually studied without including  effects of the expanding universe. However in some recent studies  black holes have been embedded in an expanding universe, in order to determine the interplay, if any, of these two dynamical processes.  Dynamical 3-space theory contains time independent solutions for black holes, which are spatial in-flows, and separately the  time dependent Hubble expansion. This theory has explained numerous puzzles in observational astrophysics and contains 3 constants; $G$, $\alpha$ - which from experimental data turns out to be the fine structure constant, and $\delta$ - which is a small but nonzero distance, possibly a Planck-type length. The Hubble expansion in the dynamical 3-space theory cannot be `switched off', forcing the study, first, of  isolated black holes coexisting with the expanding universe. It is shown that a time dependent black hole and expanding universe solution exists. The nature and implications of these solutions are discussed as they evolve over time. A dynamical network  of black holes and induced linking cosmic filaments forming  bubble structures is discussed, as a consequence of dynamical 3-space  undergoing a dynamical breakdown of homogeneity and isotropy, even in the absence of baryonic matter.}}\smallskip
\end{center}]{%

\setcounter{section}{0}
\setcounter{equation}{0}
\setcounter{figure}{0}
\setcounter{table}{0}
\setcounter{page}{1}

\markboth{D.P. Rothall and R.T. Cahill. Dynamical 3-Space: Black Holes in an Expanding Universe}{\thepage}
\markright{D.P. Rothall and R.T. Cahill. Dynamical 3-Space: Black Holes in an Expanding Universe}

\section{Introduction}
The motions of stars in galaxies are strongly affected by their central massive black holes, and that of galaxies in clusters are also affected by the expansion of the universe \cite{k1}. Then the need arises to analyse black holes in the expanding universe, with the view to checking if that expansion affects black hole characteristics. There is a long history of attempts to model this phenomenon analytically; early attempts include the Einstein-Strauss model through embedding Schwarzschild black holes in the background (FLRW) universe \cite{e1}, and also the well known McVittie solution \cite{m1}. This gradually lead to models (see \cite{g1} or \cite{c8} for overviews) which include the cosmological constant. The currently accepted work is based on theories of gravitation by Newton, and then extended by Hilbert and Einstein.  The use of these models has generated many questions about observational phenomena, such as 'supermassive' galactic central black holes \cite{g2}, bore hole anomalies \cite{a1,t1}, flat spiral galaxy rotation curves \cite{p2} and cosmic filaments \cite{v1}. The `dark matter' and `dark energy' parameters introduced are required in order to fit the Friedmann universe expansion equation to the type 1a supernovae \cite{p1,r1} and CMB data \cite{k2}. A more recent account of space and time \cite{c1} models time as a non-geometrical process (keeping space and time as separate phenomena), which leads to the dynamical 3-space theory. This theory is a uniquely determined generalisation of Newtonian Gravity (NG) expressed in terms of a velocity field, defined relative to  observers,  rather than the original gravitational acceleration field. This velocity field corresponds to a space flow, which has been detected in numerous experiments. These include gas-mode Michelson interferometer, optical fibre interferometer and coaxial cable experiments, and spacecraft Earth-flyby Doppler shift data \cite{c5}. The observational phenomena mentioned above are now gradually becoming interpreted through understanding the dynamics of space, which appears to offer an explanation for `dark matter' and `dark energy' effects \cite{c6, c7}. A brief introduction to the dynamical 3-space theory along with experimental and observational tests  is given in sects. 2-5. In sects. 6 and 7 we report the discovery of exact black hole solutions embedded in an expanding universe, and discuss the nature of their evolution over time, suggesting that primordial black holes  develop linking filaments, which in turn form a cosmic network with bubble structures.   

\section{Dynamical 3-Space}\label{sect:3space}
Process Physics \cite{c1} is a theory of reality which models time as a non-geometric process, with space-geometry and quantum physics being emergent and unified phenomena. The emergent geometry is thought of as a structured quantum-foam `space' and is found to be dynamic and fractal in nature, with its 3 dimensionality only approximate at micro scales. If non-trivial topological aspects of the quantum foam are ignored, it may be coarse-grain embedded in a 3-dimensional geometrical manifold. This embedding ultimately allows us to describe the dynamics of the quantum foam, or space, using a classical velocity field $\textit{\textbf{v}}(\textit{\textbf{r}},t)$, relative to an observer with coordinate system $\textit{\textbf{r}}$ and $t$ \cite{c6}, and  here assuming zero vorticity, $\nabla\times \textit{\textbf{v}}=\textit{\textbf{0}}$:
\vspace{0mm}
 \begin{eqnarray}
\nabla\! \cdot\!\left(\frac{\partial \textit{\textbf{v}} }{\partial t}+ (\textit{\textbf{v}}\!\cdot\! \nabla)\textit{\textbf{v}}\right)+
\frac{5 \alpha}{4}\left((tr D)^2 -tr(D^2)\right) +  \mbox{\ \ \ \ \ \ \  \ \ \ \ } \nonumber \\
\delta^2\nabla^2\left((tr D)^2 -tr(D^2)\right)+...=-4\pi G\rho,  \mbox{\  \  \  }
 D_{ij}=\frac{\partial v_i}{\partial x_j}.  \label{eqn:3space}\end{eqnarray}
where $\rho = \rho(\textit{\textbf{r}},t)$ is the usual matter density.\footnote{The $\alpha$ term in  (\ref{eqn:3space}) has been changed by a factor of ten due to a numerical error found in the analysis of borehole data. All solutions are also altered by this factor.}

The 1st term involves the Euler constituent acceleration, while the $\alpha-$ and $\delta-$ terms contain higher order derivative terms and describe the self interaction of space at different scales.
Laboratory, geophysical and astronomical data suggest that $\alpha$ is the fine structure constant $\approx 1/137$, while $\delta$ appears to be a very small but non-zero Planck-like length. The emergence of gravity  arises from the unique coupling of quantum theory to the 3-space \cite{c1a}, which determines the `gravitational' acceleration of quantum matter  as a quantum wave refraction effect,
\begin{equation}
\textit{\textbf{g}}=\displaystyle{\frac{\partial \textit{\textbf{v}}}{\partial t}}+(\textit{\textbf{v}}\cdot{\bf \nabla})\textit{\textbf{v}}+({\bf \nabla}\times\textit{\textbf{v}})\times\textit{\textbf{v}}_R
-\frac{\textit{\textbf{v}}_R}{1-\displaystyle{\frac{\textit{\textbf{v}}_R^2}{c^2}}}
\frac{1}{2}\frac{d}{dt}\left(\frac{\textit{\textbf{v}}_R^2}{c^2}\right)+...
\label{eqn:acceleration}\end{equation}  
where $\textit{\textbf{v}}_R=\textit{\textbf{v}}_0-\textit{\textbf{v}}$ is the velocity of matter relative to the local space. The 1st two terms are the Euler space acceleration, the 2nd term explains the Lense-Thirring effect when the vorticity is non-zero, and the last term explains the precession of planetary orbits.

Neglecting relativistic effects (\ref{eqn:3space}) and (\ref{eqn:acceleration}) give
\begin{equation}
\nabla\cdot\textit{\textbf{g}}=-4\pi G\rho-4\pi G \rho_{DM},
\label{eqn:E7}\end{equation}
where
\begin{eqnarray}
\rho_{DM}(\textit{\textbf{r}},t)\equiv\frac{5 \alpha}{16\pi G}\left( (tr D)^2-tr(D^2)\right)\nonumber\\ 
+\frac{\delta^2}{32\pi G}\nabla^2\left((tr D)^2-tr(D^2)\right).
\label{eqn:E7b}\end{eqnarray}
This is Newtonian gravity, but with the extra dynamical term which has been used to define an effective `dark matter' density. Here $\rho_{DM}$ is purely a space/quantum foam self interaction effect, and is  the matter density needed within Newtonian gravity to explain dynamical effects caused by the $\alpha$ and $\delta$ effects in (\ref{eqn:3space}). This effect has been shown to offer an explanation for the  `dark matter' effect in spiral galaxies,  anomalies in laboratory $G$ measurements, bore hole $g$ anomalies, and the systematics of galactic black hole masses, as noted below. When $\alpha=0$ and $\delta=0$, (\ref{eqn:E7}) reduces to Newtonian gravity.  The $\alpha-$term has the same order derivatives as the Euler term, and so cannot be neglected {\it a priori}. It was, however, missed by Newton as its consequences are not easily observable in the solar system, because of the low mass of planets relative to the massive sun. However in galaxies this term plays a major role, and the Milky Way black hole data has given evidence for that term and as well for the next higher order derivative terms.

The spatial dynamics  is non-local and instantaneous, which points to the universe being highly connected,  consistent with the deeper pre-space process physics.  Historically this was first noticed by Newton who called it action-at-a-distance. To see this  (\ref{eqn:3space}) can be written as an non-linear integro-differential equation
\begin{equation}
\frac{\partial \textit{\textbf{v}}}{\partial t}=-\nabla\left(\frac{\textit{\textbf{v}}^2}{2}\right)-G\!\!\int d^3r^\prime
\frac{\rho_{DM}(\textit{\textbf{r}}^\prime, t)+\rho(\textit{\textbf{r}}^\prime, t)}{|\textit{\textbf{r}}-\textit{\textbf{r}}^\prime|^3}(\textit{\textbf{r}}-\textit{\textbf{r}}^\prime).
\label{eqn:E8}\end{equation}
This shows a high degree of non-locality and non-linearity, and in particular that the behaviour of both $\rho_{DM}$ and $\rho$ manifest at a distance irrespective of the dynamics of the intervening space. This non-local behaviour is analogous to that in quantum systems and may offer a resolution to the horizon problem.

\section{Evidence for the $\alpha-$ and $\delta-$ dynamical terms}\label{sec:Evidence}
\subsection{$\delta = 0$ - Early Studies of Dynamical 3- Space}\label{sec:deltazero}

\begin{figure}
\hspace{2mm}\includegraphics[scale=0.26]{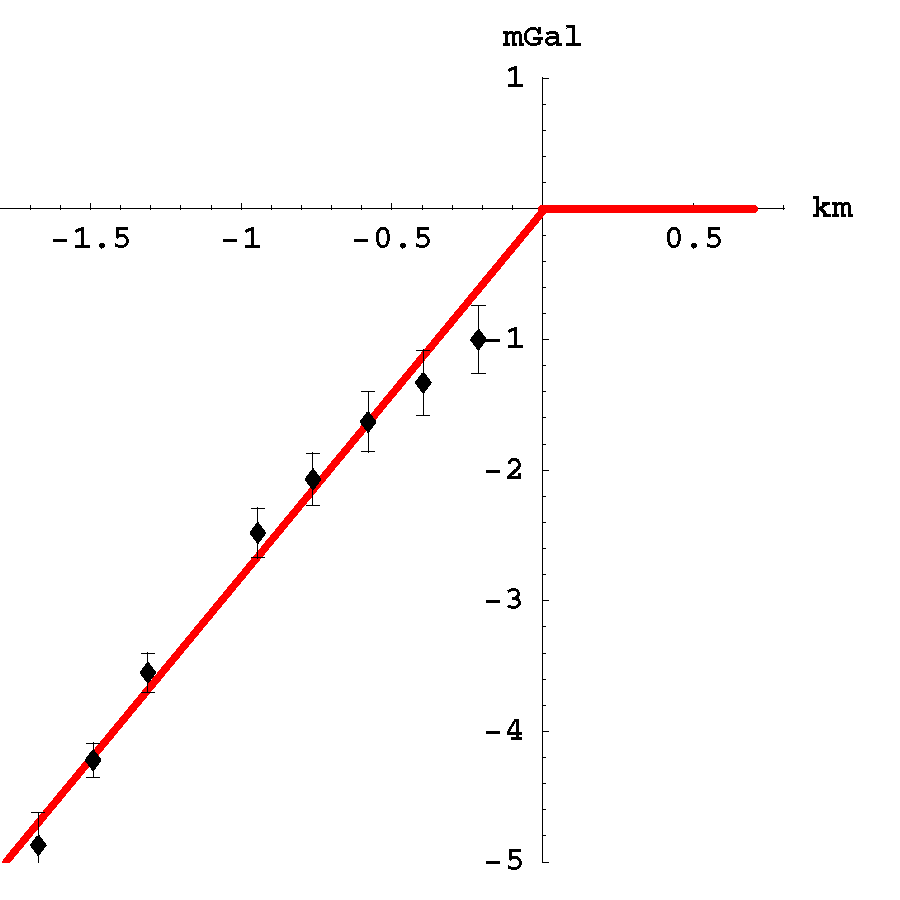}
\vspace{-4mm}\caption{\small{The Greenland ice bore hole $g$ anomaly data, giving $\alpha \approx 1/137$ from fitting the form in (\ref{eqn:borehole}). The misfit at shallow depths arises from the ice not having reached the ice-shelf full density, which is a snow compactification effect.   The Nevada rock bore hole data \cite{t1} also gives $\alpha \approx 1/137$.  The bore hole anomaly is that gravity is stronger down a bore hole than predicted by Newtonian gravity.}}
\label{fig:Greenland}\end{figure}

It has been shown that dynamical 3-space flows into matter \cite{c1a}. External to a spherically symmetric matter density $\rho(r)$, (\ref{eqn:3space}) has a time-independent radial inflow solution $v(r)  \sim  1/r^{\frac{1}{2}}$ leading to the matter inward acceleration $ g(r) \sim 1/r^2$. This happens because the $\alpha-$ and $\delta-$ dynamical terms are identically zero for this inflow speed, and explains why these significant terms were missed by Newton in explaining Kepler's Planetary Laws. However, inside a spherically symmetric mass, and in other circumstances, these terms play a significant dynamical role. Inside a spherically symmetric mass, such as the earth, Newtonian gravity and the new dynamics predict different matter accelerations,

\begin{equation}
\Delta g=g_{NG}(d)-g(d)=20\pi\alpha G \rho d+O(\alpha^2)
\label{eqn:borehole}\end{equation}
where $d<0$ is the depth. The Greenland \cite{a1} (see Fig.\ref{fig:Greenland}) and Nevada bore hole data \cite{t1}, reveal that $\alpha \approx 1/137$, the fine structure constant known from quantum theory. This suggests we are seeing a unification of gravity and the quantum theory.

In conventional theory black holes are required to have enormous quantities of actual in-fallen matter compressed into essentially a point-like region. Their $g \sim 1/r^2$ gravitational acceleration field is unable to explain flat spiral galaxy rotation curves, resulting in the invention of `dark matter'. Dynamical 3-space theory however also predicts black holes in the absence of in-fallen matter, which produce a stronger acceleration field  $g \sim 1/r$, as discussed below.  They are spherically symmetric in-flows of space, with  space not being conserved. In the absence of matter, $\rho = 0$, we set $\textit{\textbf{v}}(\textit{\textbf{r}},t) = v(r)\hat\textit{\textbf{r}}$. Previous work considered solutions of (\ref{eqn:3space}) when $\delta=0$, where the black hole solutions were found to have the  form
\begin{equation}
v(r) = -\frac{\beta}{r^\frac{5\alpha}{2}}
\label{eqn:bhdeltaeq0}\end{equation}
where $\beta$ is an arbitrary parameter for the strength of the black hole. Eqn. (\ref{eqn:3space}) also has straight-line filament solutions, with the form, when  $\delta=0$, 
\begin{equation}
v(r) = -\frac{\mu}{r^\frac{5\alpha}{4}}
\label{eqn:fildeltaneq0}\end{equation}
where $r$ is the perpendicular distance from the filament and $\mu$ is the arbitrary filament strength. The solutions (\ref{eqn:bhdeltaeq0}) and (\ref{eqn:fildeltaneq0}) contain a singularity at $r=0$ where the in-flow speed becomes infinite. Asymptotically, even when $\rho \neq 0$,   these black hole solutions  predict flat spiral galaxy rotation curves, for the inflow in (\ref{eqn:bhdeltaeq0}) gives 
$g(r)=-5\alpha \beta^2/2r^{1+5\alpha} \sim -1/r$, giving the circular orbit speed $v_0(r)=(10\alpha\beta^2)^{1/2}/2r^{5\alpha/2}$, and illustrated in Fig.\ref{fig:flat}. This suggests that the `dark matter' effect is caused by the $\alpha-$dynamical term, a space self-interaction.

\begin{figure}
\hspace{0mm}\includegraphics[scale=0.63]{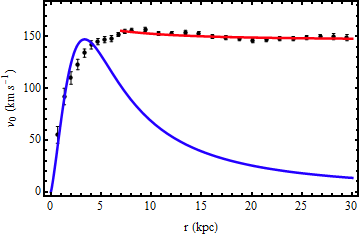}
\vspace{0mm}\caption{\small{The flat asymptotic star rotation speeds for the spiral galaxy NGC3198, with upper flat plot from the dynamical 3-space, while the lower form is from Newtonian gravity. 
The flat asymptotic form arises when $\alpha \neq 0$.}}
\label{fig:flat}\end{figure}

The Maxwell EM equations take account of the 3-space dynamics by making the change $\partial /\partial t  \rightarrow \partial /\partial t +\textit{\textbf{v}}\cdot \nabla$. Then we obtain  strong galactic light bending and lensing caused by the inflow speed in (\ref{eqn:bhdeltaeq0}), or the solar light bending when $v \sim 1/r^{\frac{1}{2}}$.
There are also recent direct experimental detections of the space flow velocity field by \cite{c5}.

\begin{figure}
\hspace{0mm}\includegraphics[scale=0.60]{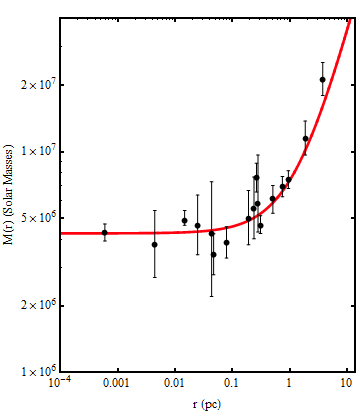}
\vspace{0mm}\caption{\small{Effective mass  data  $M(r)$ for the Milky Way SgrA* black hole, from star and gas cloud orbital data,  showing the flat regime that mimics a point-like mass, but for which there is no actual matter  contained within the black hole, and the linearly rising form beyond $r_s=$1.33pc, as predicted by  (\ref{eqn:MW}), but which is usually attributed to a constant `dark matter' density. This form is a direct consequence of the 3-space self-interactions in (\ref{eqn:3space}). The offset of the  last two points indicate the presence of actual matter.}}
\label{fig:MW}\end{figure}

\subsection{$\delta \neq 0$ - Black Holes and Filaments }\label{sec:bhfil}

More recently the $\delta \neq 0$ scenario was considered.  The form of (\ref{eqn:3space}) is expected as a semi-classical derivative expansion of an underlying quantum theory, where higher order derivatives are indicative of shorter length-scale physics. Eqn.(\ref{eqn:3space})  when $\rho = 0$ has exact two-parameter, $v_{0}$ and $\kappa \geq 1$, black hole solutions 
\begin{eqnarray}
v(r)^2 = v_{0}^2(\kappa - 1)\frac{\delta}{r}\left(1 -   \mathstrut_1\!F_1\left[-\frac{1}{2} + \frac{5\alpha}{2},-\frac{1}{2},-\frac{r^2}{\delta^2}\right]\right) \nonumber\\
- v_{0}^2\kappa\frac{8}{3}\frac{r^2}{\delta^2}\frac{\Gamma (\frac{3-5\alpha}{2})}{\Gamma (\frac{-5\alpha}{2})}\mathstrut_1\!F_1\left[1+\frac{5\alpha}{2},\frac{5}{2},-\frac{r^2}{\delta^2}\right]
\label{eqn:E2}\end{eqnarray}
where $\mathstrut_1\!F_1[a,b,w]$ is the confluent hypergeometric function. The parameters $v_{0}$ and $\kappa$ set the strength and structure of the black hole, as discussed in \cite{c6}. Eqn.(\ref{eqn:E2}) is a generalisation of  (\ref{eqn:bhdeltaeq0}), and for  $r \gg \delta$  gives
\begin{equation}
v(r)^2 \approx A\frac{\delta}{r}+B\left( \frac{\delta}{r} \right)^{5\alpha}\label{eqn:BHflowspeed}
\end{equation}
giving, from (\ref{eqn:acceleration}),  $g(r)=GM(r)/r^2$, where $M(r)$ defines an 'effective mass' contained within radius $r$, but which does not entail any actually matter,
\begin{equation}M(r)=M_0+M_0\left( \frac{r}{r_s}\right)^{1-5\alpha}\label{eqn:MW}\end{equation}
and $r_{s}$ is the distance where $M(r_{s})=2M_{0}$. This is shown in Fig.\ref{fig:MW} for the Milky Way SgrA$^*$ black hole. At large $r$ the in-flow speed becomes very slowly changing , thus predicting flat rotation curves given by \cite{c6}

\begin{equation}
v_{orb}(r)^2 = G M_{0}\left( \frac{r_{s}}{r}\right) ^{5\alpha} \frac{1}{r_{s}}.
\label{eqn:vorbital}\end{equation}

Fig.\ref{fig:BlackHoleMasses} illustrates  that for globular clusters and spherical galaxies the observational  data implies the relationship $M_{BH}=\frac{\alpha}{2}M$. Again we see that the $\alpha-$term dynamics appear to be the cause of this result, although this has yet to be derived from (\ref{eqn:3space}).
Exact filament solutions for (\ref{eqn:3space}) also exist when $\delta \neq 0$, as  a generalisation of  (\ref{eqn:fildeltaneq0}):
\begin{equation}
v(r)^2 = v_{0}^2\frac{r^2}{\delta^2}\mathstrut_1\!F_1\left[1+\frac{5\alpha}{4},2,-\frac{r^2}{2\delta^2}\right].
\label{eqn:filamentflowspeed}\end{equation}
Here $r$ is the distance perpendicular to the axis of the filament and $v(r)$ is the in-flow in that direction. The only known filament solution is for one that is infinitely long and straight. Both (\ref{eqn:E2}) and (\ref{eqn:filamentflowspeed}) are well behaved functions  which converge to  zero as $r \rightarrow 0$., i.e. the in-flow singularities are removed.

\begin{figure}
\hspace{0mm}\includegraphics[scale=0.63]{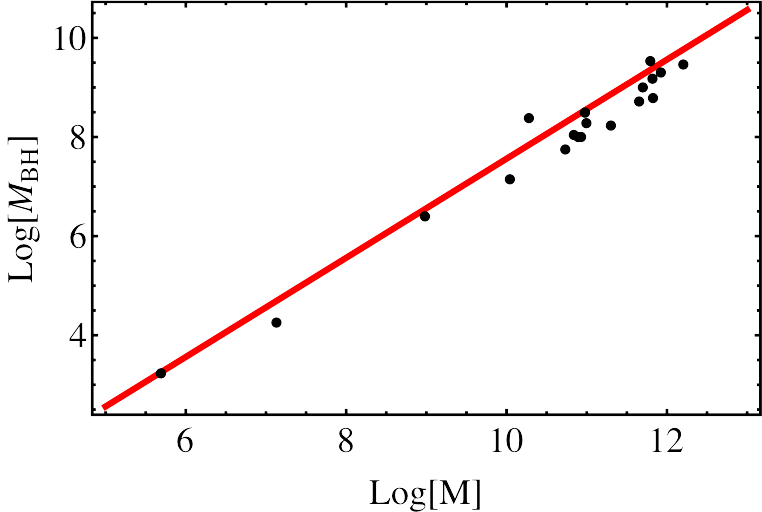}
\vspace{0mm}\caption{\small{Black hole masses $M_{BH}$ vs mass $M$, in solar masses, for the globular clusters M15 and G1, and spherical galaxies \cite{m3}. The straight line is the relation $M_{BH}=\frac{\alpha}{2}M$, where $\alpha$ is the fine structure constant $\approx$ 1/137. This demonstrates again the role of $\alpha$ in the dynamics of space and black holes.}
\label{fig:BlackHoleMasses}}\end{figure}

\section{Expanding Universe}\label{sect:expand}
Eqn.(\ref{eqn:3space})  contains a time dependent  expanding universe solution. Substituting the Hubble form  $\textit{\textbf{v}}(\textit{\textbf{r}},t)= H(t)\textit{\textbf{r}}$, and then  $H(t)$ $=\dot{a}/a$, where $a(t)$ is the universe scale factor and $\dot a(t)\equiv da(t)/dt$, we obtain
\vspace{0mm}
\begin{equation}
4a {\ddot a}+10\alpha {\dot a}^2=-\frac{16}{3}\pi G a^2 \rho
\label{eqn:Universe}\end{equation}
which is independent of $\delta$. One of the key features in (\ref{eqn:Universe}) is that even when $\rho=0$, i.e. no matter,  and $\alpha = 0$,  $\ddot{a}(t) = 0$ and  $a(t)=t/t_0$, and the universe is uniformly  increasing in scale. Here $a(t_0)=1$ and $t_0$ is the current age of the universe. This expansion of space is because the space itself is a dynamical system, and the (small) amount of actual baryonic matter merely slightly slows that expansion, as the matter dissipates space. Because of the small value of $\alpha=1/137$, the $\alpha$ term  only plays a significant role in extremely early epochs, but only if the space is completely homogeneous. In the limit $\rho \rightarrow 0$ we obtain the solution to (\ref{eqn:Universe})
\begin{eqnarray}
a(t)=\left(\frac{t}{t_0}\right)^{1/(1+5\alpha/2)} \nonumber \\
H(t)=\frac{1}{t(1+5\alpha/2)}.
\label{eqn:Hubble}\end{eqnarray}
which, as also reasoned by \cite{m2}, predicts the emergence of a uniformly expanding universe after neglecting the $\alpha$ term.  This allows a  fit to the type 1a supernovae magnitude-redshift data (Fig.\ref{fig:HubbleSN}), as discussed in \cite{c7}, and suggests that the dynamical 3-space theory also offers an explanation for the `dark energy' effect. The $\Lambda$CDM parameters $\Omega_\Lambda=0.73, \Omega_M=0.27$, follow from either fitting to the supernovae data, or equally well, fitting to the uniformly expanding universe solution in  (\ref{eqn:Hubble}) \cite{c7}.
Via the dynamical 3-space solution the supernovae data gives an age for the universe of $t_0 = 13.7$ Gy.

\begin{figure}
\hspace{0mm}\includegraphics[scale=0.47]{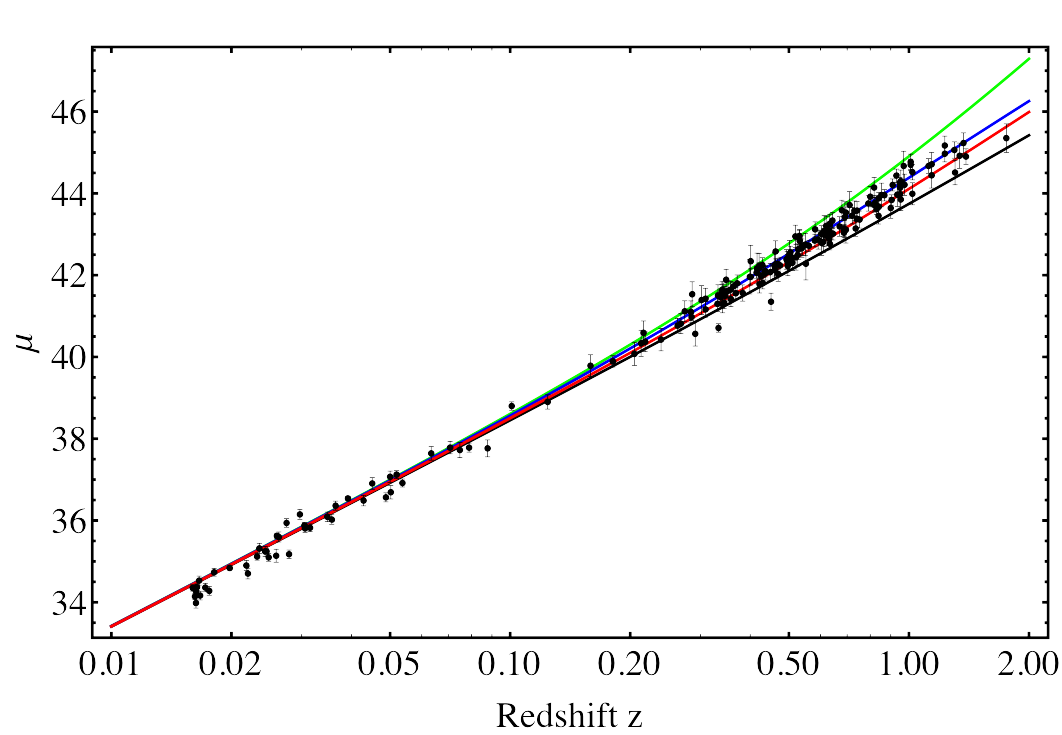}
\vspace{0mm}\caption{\small{Supernovae magnitude-redshift data. Upper curve (light blue) is `dark energy' only $\Omega_\Lambda=1$. Next curve (blue) is best fit of `dark energy'-`dark-matter' $\Omega_\Lambda = 0.73$. Lowest curve (black) is `dark matter' only $\Omega_\Lambda = 0$. 2nd lowest curve (red) is the uniformly expanding universe, and also predicted by dynamical 3-space (\ref{eqn:Hubble}).
}}
\label{fig:HubbleSN}\end{figure}

\section{Black Hole - Expanding Universe }\label{sect:perturb}
The Hubble solution (\ref{eqn:Hubble}) does not contain a free parameter, i.e. in the dynamical 3-space theory the universe necessarily expands, and hence it cannot be ignored when considering black holes and filaments. Since any radially flowing and time dependent $v(r,t)$ (i.e. containing both outflows and inflows) has spherical symmetry, (\ref{eqn:3space}) becomes, in the absence of matter
\begin{eqnarray}
\frac{\partial}{\partial t}\left(\frac{2v}{r}+v'\right) +v v''+2\frac{v v'}{r}+\left(v'\right)^2+\frac{5\alpha}{2}\left(\frac{v^2}{r^2}+\frac{2vv'}{r}\right)+ \nonumber \\
+\frac{\delta^2}{4r^4}\left(2v^2+2r^2(v')^2+6r^3v'v''\right)\mbox{\ \ \ \ \ \ \  \ \ \ \ \ \ \ \ \ \ \ \ \ \   } \nonumber \\
+\frac{\delta^2}{4r^4}\left(-4rvv'+2r^2vv''+2r^3vv'''\right)=0 \mbox{\ \ \ \ \ \ \  \ \ \ \  }
\label{eqn:A1}\end{eqnarray}
where $v^\prime \equiv \partial v/\partial r$. Now consider the black hole - expanding universe  ansatz
\begin{equation}
\textit{\textbf{v}}(\textit{\textbf{r}},t) =H(t)\textit{\textbf{r}}+ w(r,t)\hat{\textit{\textbf{r}}}
\label{eqn:vhubpert}\end{equation}
where $w(r,t)$ is the spherically symmetric black hole inflow. After substituting this form we obtain a time dependent  equation for $w(r,t)$.
However by  setting $w(r,t)=R(r)/t$  this time dependence is resolved, and (\ref{eqn:A1}) now may be solved for $R(r)$, implying that the Hubble outflow and black hole inflow are inseparable and compatible phenomena. 
Asymptotically, when  $r \gg \delta$, the resulting equation for $R(r)$  has the solution
\begin{equation}
R(r) = -\frac{\nu}{r^\frac{5\alpha}{2}}, \mbox{\ \ \   and so \ \ } w(r,t)= -\frac{\nu}{r^{\frac{5\alpha}{2}}t}
\label{eqn:hubbhdeltaeq0}\end{equation}
which is the original black hole solution (\ref{eqn:bhdeltaeq0}), but now with an inverse time dependence. Eqn (\ref{eqn:vhubpert}) is for the black hole located at $\textit{\textbf{r}}=\textit{\textbf{0}}$. For a black hole comoving  with the local Hubble space flow the solution of (\ref{eqn:3space}) is
\begin{equation}
\textit{\textbf{v}}(\textit{\textbf{r}},t) =H(t)\bf{r}^\prime+ w(r',t)\bf{\hat r}^\prime
\label{eqn:vhubpertmov}\end{equation}
where $\bf{r}^\prime =\textit{\textbf{r}}-a(t)\textit{\textbf{r}}_{BH}$  when the observer is at  $\textit{\textbf{r}}=\textit{\textbf{0}}$, and the black hole is located at $a(t)\textit{\textbf{r}}_{BH}$. Macroscopic black holes are expected to form from coalescence of mini primordial black holes. 

A consequence of (\ref{eqn:vhubpert}) is that for any black hole there exists a critical radius $r_{c}$ where the spatial inflow into the black hole is equal and opposite to the Hubble expansion, Fig.\ref{fig:hubblepert}, so defining a sphere of influence. Test particles placed inside $r_{c}$ are attracted to the black hole due to gravity, while those placed outside $r_{c}$, and at rest wrt the local space, recede from it due to expansion.
This critical radius is found to remain independent of time, i.e. $r_{c}$ only depends on the black hole strength $\nu$. $r_{c}$ is expected to be sufficiently large that the black hole - star distance $r$ in a galaxy today is negligible compared to $r_{c}$, i.e. $r \ll r_{c}$, therefore not affecting the size of the galaxies themselves. This effect would more likely be evident at a distance which galaxies are separated by, as suggested by the galaxy cluster data in \cite{n1}. For a Hubble constant \mbox{$H_0 = 74$ km s$^{-1}$ Mpc$^{-1}$}, and using (\ref{eqn:vorbital}) for the in-flow speed, solving for \mbox{$v_{orb}(r_{c})=H_{0}r_{c}$}  for the Milky Way SgrA$^*$ black hole data (fig.\ref{fig:MW}) yields \mbox{$r_{c}=1.6$ Mpc}. For multiple black holes  in the expanding space, (\ref{eqn:3space}) implies a more complex time evolution.

\begin{figure}
\centering
\includegraphics[scale=0.6]{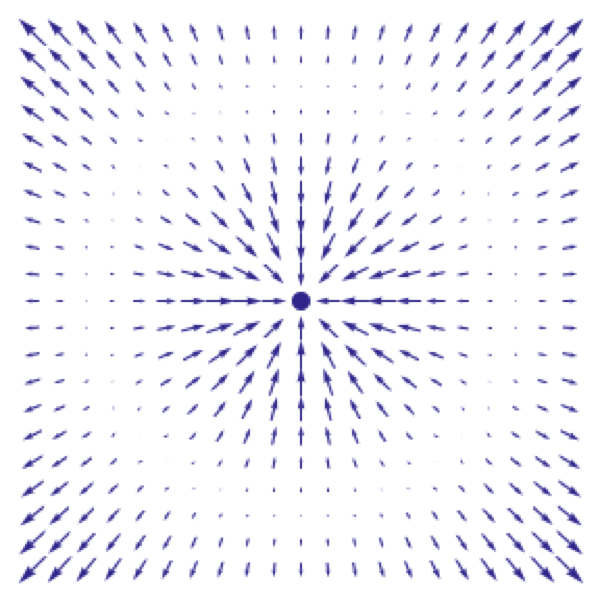}
\vspace{0mm}\caption{\small{Schematic 3-space  velocity for an isolated black hole embedded in an expanding universe, see (\ref{eqn:vhubpert}), showing radius at which flow reverses, defining the black holes sphere of influence}.}
\label{fig:hubblepert}\end{figure}

\begin{figure}
\hspace{10mm}\includegraphics[scale=0.6]{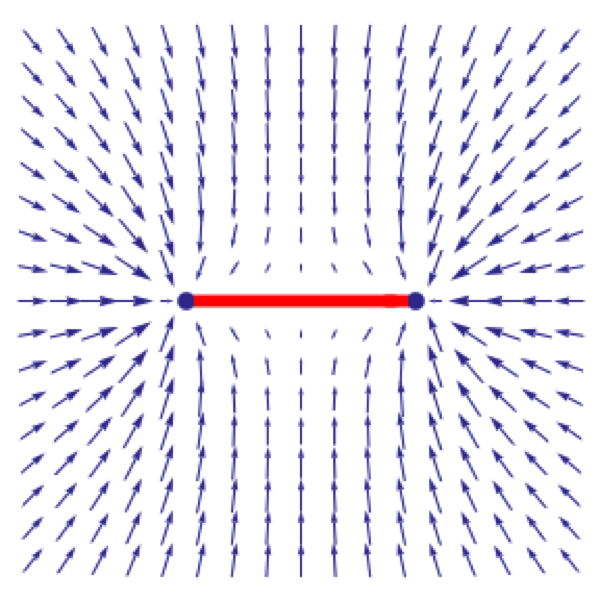}
\vspace{0mm}\caption{\small{3-space in-flow velocity for two black holes located within their spheres of influence. Note the emergence of a filament forming between the black holes, indicative of a black-hole - filament network formation, see fig.\ref{fig:bubble}.}
\label{fig:2bh}}\end{figure}

\begin{figure}
\vspace{-3mm}\hspace{5mm}\includegraphics[scale=0.7]{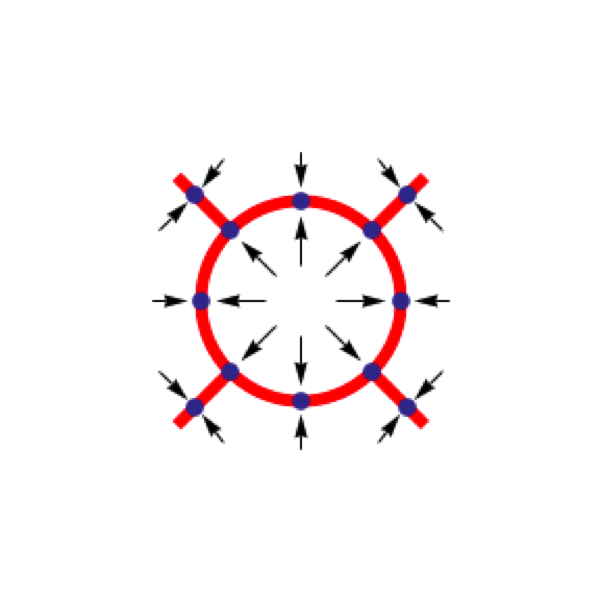}
\vspace{-13mm}\caption{\small{2D schematic section of a cosmic network of black holes and induced filaments. Vectors indicate 3-space flow, both within the bubble from the Hubble space expansion, and inwards to black holes (dots) and filaments (red lines).  Only this bubble structure, shown here in cross-section, appears to be stable wrt the Hubble expansion.}
\label{fig:bubble}}\end{figure}

\section{Induced Filaments and Bubble Networks}
We have seen that the dynamical 3-space theory offers possible explanations for many phenomena, including that of an isolated black hole coexisting with the Hubble expansion. It also has filament solutions, in the absence of the Hubble expansion. However with multiple black holes a new feature appears to emerge, namely cosmic networks of black holes and induced filaments.  First note that the black hole inflow speed  in (\ref{eqn:BHflowspeed}) is essentially very long range, resulting in the matter acceleration $g(r)\sim -1/r$, which is a key feature of these black holes, and may explain the `dark matter' effect. However this long range in-flow raises the question of how multiple black holes coexist when located within one another's sphere of influence?  Fig.\ref{fig:2bh} shows the vector addition of the inflows for two black holes. This cannot be a solution of (\ref{eqn:3space}) as it is non-linear and so does not have a superposition property. Whence this flow must evolve over time. Indeed the evolving flow appears to form a filament  connecting the two black holes.  However even then there remains a long range inflow, which would lead to further filaments connecting black holes within their range of influence.  These black holes are remnants of the early formation of space, and imply that (\ref{eqn:3space}) will undergo a dynamical breaking of symmetry, from an essentially homogeneous and isotropic  3-space, to a network of black holes and induced filaments.  Note that the matter content of the universe is very small, and does not play a key role in this structure formation. A possible dynamically stable  3-space structure is shown in fig.\ref{fig:bubble}, which entails this network forming a bubble structure with the network defining a `surface' for the bubbles. The stability of this is suggested by noting that  the Hubble expansion within the interior of each bubble is now consistent with the inflow into the black holes and filaments, and so there is no longer a dynamical clash between the long range flows. Bubble structures like these are indeed found in the universe, where galaxies are observed to be joined by filaments  lying on spherical surfaces, filled with large voids \cite{d1,r2}.

\section{Conclusions}\label{sect:conclusions}
It is clear that instead of studying black-hole only cases, we need to model astrophysical and cosmological phenomena  embedded in an expanding universe. The dynamical 3-space theory naturally forces us to do this, as there is no free parameter to switch off the emergent expanding universe solution, and so must be included. It has been shown that the long range  black hole solutions found previously hold while embedded in an expanding universe. It is suggested that the time dependent nature of these new solutions explains in part the observed cosmic web.   It appears that the dynamics of the 3-space, in the presence of primordial black holes, essentially defects in the  space emerging from the quantum foam, renders a homogeneous and isotropic universe dynamically unstable, even without the presence of matter, resulting in a spatial bubble network. The long range $g \sim 1/r$ of both the black holes and induced filaments will cause matter to rapidly infall  and concentrate around these spatial structures, resulting in the precocious formation of galaxies.


\begin{thebibliography}{99}
\bibitem{a1} Ander M.E., Zumberge M.A., Lautzenhiser T., Parker R.L., Aiken C.L.V., Gorman M.R., Nieto M.M., Cooper A.P.R., Ferguson J.F., Fisher E., McMechan G.A., Sasagawa G., Stevenson J.M., Backus G., Chave A.D., Greer J., Hammer P., Hansen B.L., Hildebrand J.A., Kelty J.R., Sidles C. and Wirtz J.,  Test of Nevvton's Inverse-Square Law in the Greenland Ice Cap, {\it Phys. Rev. Lett} 1989, v.62, 985.

\bibitem{c1} Cahill R.T., Process Physics: From Information Theory to Quantum Space and Matter, {\it Nova Science Publishers, New York}, 2005.

\bibitem{c1a} Cahill R.T.,  {\it  Dynamical Fractal 3-Space and the Generalised Schr\"{o}dinger Equation: Equivalence Principle and Vorticity Effects}, {\it Progress in Physics}, 2006, 1, 27.

\bibitem{c2} Cahill R.T.,  Dynamical 3-Space: Cosmic Filaments, Sheets and Voids, {\it Progress in Physics},  2011, 2, 44.

\bibitem{c5} Cahill R.T.,   Charactersiation of Low Frequency Gravitational Waves from Dual RF Coaxial-Cable Detector: Fractal Textured Dynamical 3-Space, {\it Progress in Physics},  2012, 3, 3.

\bibitem{c6} Cahill R.T. and Kerrigan D.,  Dynamical Space: Supermassive Galactic Black Holes and Cosmic Filaments,  {\it Progress in Physics}, 2011,  v.4,  50.  

 \bibitem{c7} Cahill R.T. and  Rothall D.,  Discovery of Uniformly Expanding Universe, {\it Progress in Physics}, 2012, 4, 65.

\bibitem{c8} Carr B.J., in Sanz J.L. and  Goicoechea L.J., eds.,  Observational and Theoretical Aspects of Relativistic Astrophysics and Cosmology, {\it World Scientific, Singapore}, 1985.

\bibitem{d1} De Lapparent V., Geller M.J., Huchra J.P., A Slice of the Universe, {\it  ApJ}, 1986, 302, L1.

\bibitem{e1} Einstein A. and  Straus E.G.,   The Influence of the Expansion of Space on the Gravitation Fields Surrounding the Individual Stars, {\it  Rev. Mod. Phys.}, 1945, 17, 120.

\bibitem{g2} Ghez A.M., Klein B.L., Morris M., Becklin E.E.,  High Proper Motions in the Vicinity of SgrA$^*$: Unambiguous Evidence of a Massive Central Black Hole ,  {\it  ApJ}, 1998, 509.

\bibitem{g1} Gibbons G.W. and Maeda K.,  Black Holes in an Expanding Universe, {\it Phys. Rev. Lett},  2010, v.104, 131101.

\bibitem{k1}   Karachentsev, I.D., Dolphin, A.E., Geisler, D., Grebel, E.K., Guhathakurta, P.,
Hodge, P.W., Karachentseva, V.E., Sarajedini, A., Seitzer, P., Sharina, M.E., The M81 Group of Galaxies: New Distances, Kinematics
and Structure,   A\&A, 2002, 383, 125.

\bibitem{k2} Komatsu E.,  Smith K. M.,  Dunkley J.,  Bennett C. L.,  Gold B.,  Hinshaw G.,  Jarosik N.,  Larson D.,  Nolta M. R.,  Page L.,  Spergel D. N.,  Halpern M. ,  Hill R. S., Kogut A.,  Limon M., Meyer  S. S. ,  Odegard N. , Tucker G. S. ,  Weiland J. L. ,  Wollack E.,  Wright E. L., 
  Seven-Year Wilkinson Microwave Anisotropy Probe (WMAP) Observations: Cosmological Interpretation,  {\it ApJS}, 2010, 192, 18.

\bibitem{m3} Marconi A. and  Hunt L.K.,    The Relation Between Black Hole Mass, Bulge Mass, and Near-Infrared Luminosity, {\it  ApJL},  2003, 589, L21.

\bibitem{m1} McVittie G.C.,   The Mass-Particle in an Expanding Universe, {\it  MNRAS},  1933, 93, 325.

\bibitem{m2}  Melia F.   and  Shevchuk A.S.H.,  The $R_h=ct$ Universe, {\it MNRAS}, 2012, v. 419, 3, 2579.


\bibitem{n1} Nandra R., Lasenby A.N., Hobson M.P.,   The Effect of an Expanding Universe on Massive Objects,   {\it MNRAS}, 2012, 422, 2945.


\bibitem{p1} Perlmutter S., Aldering G., Goldhaber G., Knop R.A., Nugent P.,  Castro P.G., Deustua S., Fabbro S., Goobar A., Groom D.E.,  Hook I.M., Kim A.G., Kim M.Y., Lee J.C.,  Nunes N.J., Pain R.,  Pennypacker C.R., Quimby R., Lidman C., Ellis R.S., Irwin M., McMahon R.G., Ruiz-Lapuente P., Walton N., Schaefer B., Boyle B.J., Filippenko A.V., Matheson T., Fruchter A.S.,  Panagia N., Newberg H.J.M. and  Couch W.J., Measurement of $\Omega$ and $\Lambda$ from 42 High-Redshift Supernovae, {\it Astrophys.J.}, 1999, v.517, 565.

\bibitem{p2}Persic M., Salucci P. and Stel F.  The Universal Rotation Curve of Spiral Galaxies: I. The Dark Matter Connection, {\it MNRAS},1996, v.281, No.1, 27.


\bibitem{r2} Ratcliffe A., Shanks T., Broadbent A., Parker Q.A., Watson F.G., Oates A.P., Fong R., Collins C.A., 1996, The Durham/UKST Galaxy Redshift Survey - I. Large-Scale Structure in the Universe,  {\it MNRAS}, 281, L47.

\bibitem{r1} Riess A.G.,  Filippenko A.V.,  Challis P.,  Clocchiattia A.,  Diercks A.,  Garnavich P.M.,  Gilliland R.L., Hogan C.J.,  Jha S.,  Kirshne R.P., Leibundgut B., Phillips M.M., Reiss D.,  Schmidt B.P.,  Schommer R.A.,  Smith R.C.,  Spyromilio J.,  Stubbs C.,  Suntzeff N.B. and   Tonry J.,  Observational Evidence from Supernovae  for an Accelerating Universe and Cosmological Constant, {\it Astron.J.}, 1998,  v.116,  1009.



\bibitem{t1} Thomas J. and  Vogel P., Testing the Inverse-Square Law of Gravity in Boreholes at the Nevada Test Site,  {\it Phys. Rev. Lett.},  1990, 65, 1173.

\bibitem{v1} Vachaspati T.,   Cosmic Strings and the Large-Scale Structure of the Universe,  {\it Phys. Rev. Lett.},  1986, 57, 1655.

\end{thebibliography}
\end{document}